\documentclass[a4paper,fleqn]{cas-dc}

\usepackage[utf8]{inputenc}
\usepackage[T1]{fontenc}
\usepackage{amsmath,amssymb,amsfonts,amsthm}
\usepackage{graphicx}
\usepackage{blindtext}
\usepackage{textcomp}
\usepackage{braket}
\usepackage{hyperref}
\usepackage{subcaption}
\usepackage{float}
\usepackage{tikz}
\usepackage{quantikz}
\usepackage{pgfplots}
\usepackage{xfrac}
\usepackage{appendix}
\usepackage{tabularx}
\usepackage{listings}
\usepgfplotslibrary{groupplots}

\usepackage{enumitem}
\usepackage{xstring}
\usepackage{pgfplotstable}
\usepackage{booktabs,multirow}
\usetikzlibrary{arrows.meta, positioning, fit, backgrounds, shapes.geometric, calc}
\pgfplotsset{compat=1.18}
\usepgfplotslibrary{groupplots}

\usepackage{hyperref}
\hypersetup{
    colorlinks = true,
    linkcolor = blue!50!black,
    citecolor = blue!50!black,
    filecolor = blue!50!black,
    urlcolor = blue!50!black,
}

\definecolor{qprimary}{HTML}{334261}    
\definecolor{qquantum}{HTML}{8250DF}    
\definecolor{qclassic}{HTML}{D9B112}    
\definecolor{qentangle}{HTML}{9527A3}   
\definecolor{qproduct}{HTML}{0969DA}    
\definecolor{qmeasured}{HTML}{8C959F}

\definecolor{clive}{HTML}{0072B2}       
\definecolor{cent}{HTML}{E69F00}        
\definecolor{cmeas}{HTML}{009E73}       

\lstdefinestyle{python}{
    language=Python,
    basicstyle=\ttfamily\small,
    keywordstyle=\color{blue}\bfseries,
    commentstyle=\color{gray}\itshape,
    stringstyle=\color{teal},
    numberstyle=\tiny\color{gray},
    numbers=left,
    frame=single,
    breaklines=true,
    showstringspaces=false
}
\usepackage[numbers,sort&compress]{natbib}

\def\tsc#1{\csdef{#1}{\textsc{\lowercase{#1}}\xspace}}
\tsc{WGM}
\tsc{QE}

\begin{document}
\let\WriteBookmarks\relax
\def\floatpagepagefraction{1}
\def\textpagefraction{.001}

\shorttitle{
  Q2NSViz: An Open-source Standalone Visualizer for Quantum Network Simulations
}    

\shortauthors{
  F.~Mazza, M.~Caleffi, A.~S.~Cacciapuoti
}  

\title [mode = title]{
  Q2NSViz: An Open-source Standalone Visualizer for Quantum Network Simulations
}  

\author{Francesco Mazza} 
\author{Marcello Caleffi} 
\author{Angela Sara Cacciapuoti}

\cormark[1]

\ead{angelasara.cacciapuoti@unina.it}

\affiliation{
    organization={Quantum Internet Research Group, University of Naples Federico II},
    city={Naples},
    postcode={80125}, 
    country={Italy}
}

\begin{abstract}
  The unique and non-classical features of quantum networks make their simulation and intuitive understanding inherently difficult. In this work, we present Q2NSViz, an open-source Python-based visualization tool for replaying and inspecting quantum-network simulation traces. Q2NSViz replays the time evolution of the simulated network as recorded in the trace: physical topology, stored and in-flight qubits, classical bits and packets, measurements, and entanglement relationships. In this way, it exposes not only physical connectivity, but also the dynamic entanglement-induced structure produced, consumed, and transformed by protocol execution.
  Q2NSViz is built around a decoupled JSON/NDJSON trace contract, a Qt-free replay engine, and an interactive PyQt6 interface, making it a standalone companion to Q2NS and reusable by other simulation backends that emit the same trace format. By turning execution traces into navigable and reproducible visual artifacts, Q2NSViz provides a zero-coding tool for researchers and educators, narrowing the gap between abstract protocol logic and concrete execution.
\end{abstract}


\maketitle

\setlength{\textfloatsep}{8pt plus 2pt minus 2pt}
\setlength{\intextsep}{8pt plus 2pt minus 2pt}

\begingroup
  \footnotesize
  \setlength{\tabcolsep}{4pt}
  \renewcommand{\arraystretch}{1.05}
  \noindent
  \begin{tabularx}{\columnwidth}{@{}>{\raggedright\arraybackslash}p{0.40\columnwidth} >{\raggedright\arraybackslash}X@{}}
    \bfseries\normalfont Metadata & \bfseries\normalfont Description \\
    \midrule
    Current code version            & v0.2.0 \\
    Permanent link to repository    & \url{github.com/QuantumInternet-it/q2nsviz} \\
    Permanent archive (DOI)         & \url{doi.org/10.5281/zenodo.21216676} \\
    Legal code license              & MIT \\
    Code versioning system          & git \\
    Languages, tools, services      & Python~3, PyQt6, Matplotlib, Q2NS \\
    Compilation requirements, dependencies & Python~$\geq$3.12, PyQt6~$\geq$6.6, Matplotlib~$\geq$3.7 \\
    Developer documentation         & Repository README, trace-format specification, Doxygen API, bundled example traces \\
    Support email                   & angelasara.cacciapuoti@unina.it \\
    \bottomrule
  \end{tabularx}
\endgroup
\medskip

\section{Motivation and Significance}
Quantum networks differ from classical networks because their operation depends on distributed entanglement resources that are stateful and consumed by protocol execution~\cite{Kim-08}. Entanglement induces a connectivity graph that may diverge from the physical network graph, and that evolves over time as protocols generate, transform, consume, or discard entangled states~\cite{CalCac-26,CacCal-26}. A meaningful view of the execution must track both classical network events and the lifecycle of the qubits, including the evolving entanglement structure~\cite{PeaMazCal-26,MazPeaCal-26}. State-of-the-art quantum-network simulators such as \textit{Q2NS} provide the execution substrate for these studies, enabling protocol modeling, performance evaluation, and analysis under different physical and architectural conditions~\cite{coopmans2021netsquid, wu2021sequence, KimBacBha-25, PeaMazCal-26}. 
Yet the output of such simulations often remains difficult to inspect: the relevant dynamics are distributed across low-level logs, diagnostics, and trace files. In particular, the entanglement structure produced by the protocol is not directly visible in the raw output, even though it is central to understanding the simulated execution.

In this work, we present Q2NSViz, an open-source standalone Python tool for replaying and inspecting quantum-network traces. Q2NSViz provides a time-resolved view of the simulated execution, combining the physical topology with the location and lifecycle of qubits, classical communication, measurements, and entanglement-induced connectivity. It allows researchers and educators to load, replay, and inspect traces without writing visualization code. Q2NSViz builds on the trace-replay concept introduced by the browser-based viewer distributed with Q2NS~\cite{PeaMazCal-26}, but re-engineers its realization as an independent Python package. In particular, the package is organized around the replay engine detailed in Sec.~2, which is decoupled from the graphical interface implemented in PyQt6. Q2NSViz is released under the MIT license. 

Although Q2NS is the reference producer, Q2NSViz relies on a decoupled JSON trace contract: any simulator or script emitting the same format can generate traces that Q2NSViz can replay. Unlike the experiment-building Virtual Lab~\cite{MigJanGra-22} and the simulator-specific visualization or configuration interfaces associated with QuantumSavory~\cite{KimBacBha-25} and SeQUeNCe~\cite{wu2021sequence}, Q2NSViz replays completed traces independently of the producing simulator, exposing the temporal evolution of qubits, entanglement-induced connectivity, and classical communication. This separation does not prevent integration with other frontends, which can use the documented trace format or, for Python-based applications, directly import the replay engine independently of its graphical interface.

\begin{figure*}[t]
  \centering
  \includegraphics[width=0.95\textwidth]{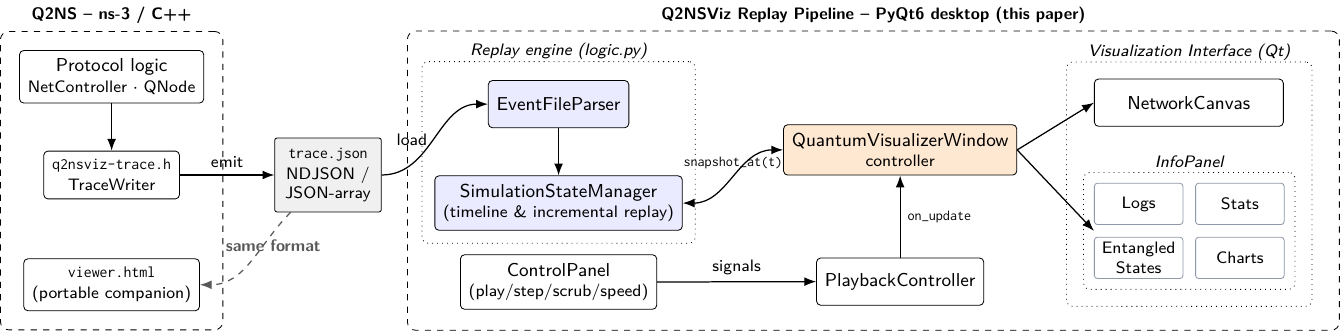}
  \caption{Q2NSViz software architecture and replay pipeline. The simulator and visualizer are decoupled through a typed NDJSON/JSON-array trace format that feeds both the desktop application and the portable browser companion shipped with Q2NS~\cite{PeaMazCal-26}. The Qt-free replay engine (\texttt{logic.py}) processes the trace and exposes a programmatic API independent of the graphical interface.}
  \label{fig:arch}
\end{figure*}

\section{Software Description}
Q2NSViz is organized as a trace-driven replay workflow rather than as a monolithic visualizer. Fig.~\ref{fig:arch} illustrates the architecture and its core concepts: a decoupled JSON/NDJSON trace format, which acts as the contract between simulator and viewer; a Qt-free replay engine, which parses the trace and replays the network state it records; and a PyQt6 desktop interface, which renders this state through interactive views. Keeping the replay logic independent of the frontend lets traces from Q2NS, or from any compatible producer, be inspected through the same execution path, and lets the interface be updated while leaving the replay engine unchanged.

\begin{figure*}[t]
  \centering
  \includegraphics[width=0.96\textwidth]{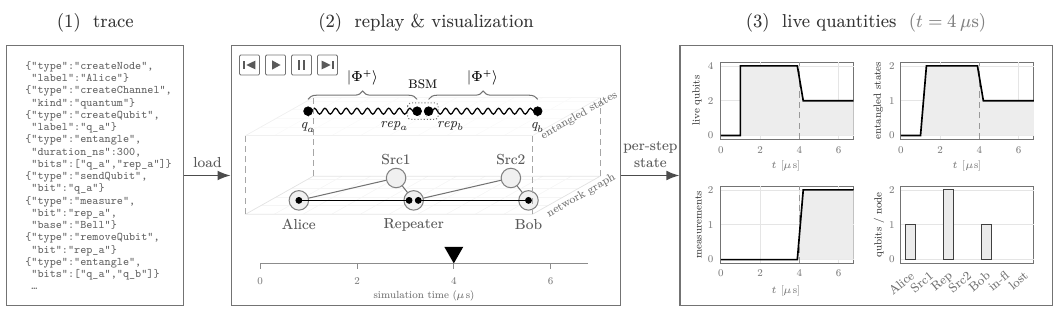}
  \caption{Q2NSViz logical representation of the trace-replay and visualization workflow.  Simulation events~(1) are replayed to obtain separate logical views of the physical network graph and the entanglement-induced connectivity graph~(2) -- capturing links that transcend physical proximity -- alongside a navigation timeline. In the graphical interface, these views are overlaid on the same canvas. Panel~(3) illustrates the time evolution of \emph{live quantities}: active qubits, entangled states, and measurements. At $t\approx4\,\mu$s an entanglement swap is in progress: the repeater still holds the two memory qubits of the Bell pairs $\ket{\Phi^+}$, both under measurement.}
  \label{fig:overview}
\end{figure*}

\begin{figure}
  \centering
  \includegraphics[width=\columnwidth]{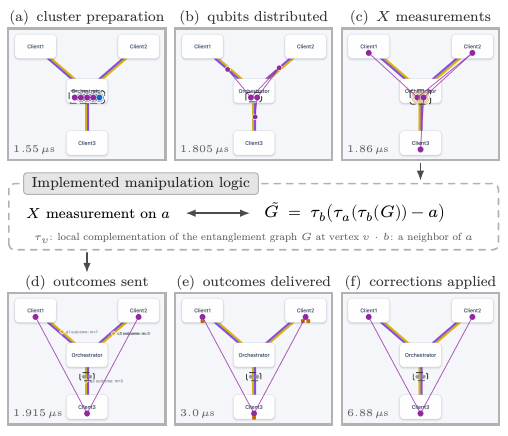}
  \caption{Graphical views generated by the Q2NSViz interface during graph-state manipulation (live-metric plots omitted for readability). A five-qubit linear cluster state is prepared~(a) and distributed to three clients~(b); two $X$ measurements are replayed through the reported graph-transformation rule~(c); the measurement outcomes are sent~(d), delivered~(e) and the corresponding corrections applied~(f). Dashed boxes denote the orchestrator qubits.}
  \label{fig:graphstate}
\end{figure}

Fig.~\ref{fig:overview} shows the replay workflow end-to-end on a repeater-swap trace, in which the Bell-state measurement at the repeater transfers the entanglement to the end qubits $q_a$ and $q_b$. The logical view deliberately separates the physical network graph, traditionally visualized by classical network animator tools such as NetAnim~\cite{ns3}, from the entanglement-induced connectivity graph, which has no classical counterpart and is the distinctive feature of Q2NSViz; the graphical interface overlays the two on the same canvas (Fig.~\ref{fig:graphstate}).

The main entities involved in the replay workflow are:
\begin{itemize}
  \setlength{\itemsep}{0pt}\setlength{\parskip}{0pt}
  \item[i.] \texttt{EventFileParser}: ingests the trace file and parses its events into the structured representation.
  \item[ii.] \texttt{SimulationStateManager}: builds a keyframe timeline of parsed events and replays them incrementally, against sparse checkpoints, tracking qubits, cbits, and entanglement relations at any requested simulation time.
  \item[iii.] \texttt{QuantumVisualizerWindow}: acts as the controller between replay engine and interface, advancing the clock in response to \texttt{ControlPanel} (user) signals, querying \texttt{snapshot\_at(t)}, and pushing the resulting state to the visualization components.
  \item[iv.] \texttt{Views}: render the current state through the network canvas and the tabbed information panel, including event logs, statistics, entanglement-state views, and charts.
\end{itemize}

\begin{table}[t]
  \centering
  \footnotesize
  \setlength{\tabcolsep}{2.2pt}
  \begin{tabular*}{\columnwidth}{@{\extracolsep{\fill}}rrrrrrr@{}}
    \toprule
    & \multicolumn{2}{c}{\bfseries Per trace} & \multicolumn{2}{c}{\bfseries \texttt{snapshot\_at(t)}} & \multicolumn{2}{c}{\bfseries Window} \\
    \cmidrule(lr){2-3} \cmidrule(lr){4-5} \cmidrule(l){6-7}
    \bfseries Events & \bfseries Loading & \bfseries Memory & \bfseries Next & \bfseries Random & \bfseries Step & \bfseries Jump \\
    \midrule
    $10^3$ & 3.7~ms & 19~MB  & 13 $\mu$s & 49 $\mu$s & 1.6~ms & 2.0~ms \\
    $10^4$ & 32~ms  & 31~MB  & 13 $\mu$s & 65 $\mu$s & 1.7~ms & 2.7~ms \\
    $10^5$ & 0.43~s & 155~MB & 15 $\mu$s & 71 $\mu$s & 2.1~ms & 3.1~ms \\
    $10^6$ & 6.0~s  & 1.4~GB & 16 $\mu$s & 78 $\mu$s & 2.4~ms & 9.6~ms \\
    \bottomrule
  \end{tabular*}
  \caption{\raggedright Median performance (five runs, three at $10^6$ events; Apple M2 laptop) versus trace length on synthetic repeater-chain traces: loading and peak memory; \texttt{snapshot\_at(t)} at the next keyframe and at a random instant; update of the whole window at a playback step and at a jump.}
  \label{tab:perf}
\end{table}

While loading a trace, the replay engine compiles its events into atomic state changes, or actions, and replays them once: it records what only accumulates along the trace, such as the qubits ever created or measured, and it stores a checkpoint about every 256 actions, holding only the state alive at its time. Hence, \texttt{snapshot\_at(t)} restores the nearest checkpoint before \texttt{t}, found by binary search, and replays at most about 256 actions, i.e., its cost grows with the live state and with the interval between checkpoints, and only logarithmically with the length of the trace. Table~\ref{tab:perf} reports the resulting performance on synthetic traces of up to one million events: as the trace grows a thousandfold, the latency of reaching an arbitrary instant grows less than twofold, whereas a replay from scratch grows with the length of the trace, and a playback step of the whole window stays within the per-frame budget of a 60~Hz refresh rate. On $10^5$ events, the checkpoints reduce the latency at a random instant by more than two orders of magnitude, with no measurable cost in memory.

Transport controls (play/step/scrub/speed) let users navigate the timeline and observe the network's evolution.
The network canvas renders network nodes, stored and in-flight qubits and their color-coded status, channel types, and entanglement-induced connectivity. Entanglement is shown through an \textit{entanglement graph} where qubits are vertices and their entanglement relationships are edges. This view exposes the entanglement connectivity structure and the effects of Bell-state measurements (BSM) and graph-state Pauli measurements via local complementation \cite{HeiEisBri-04, PeaMazCal-26, MazPeaCal-26}. Q2NSViz does not simulate or reconstruct quantum states: it visualizes the state and entanglement metadata contained in the trace, and it updates the entanglement graph through graph operations only.

Q2NSViz currently supports thirteen trace event types, summarized in Table~\ref{tab:events}. The \texttt{entangle}, \texttt{measure} and \texttt{graphMeasure} events also accept an optional \texttt{duration\_ns} field that models the latency of a state transition (e.g. unentangled to entangled or alive to measured), shown as a gating ring during the \texttt{duration\_ns} window that precedes the event timestamp. On load, the engine also validates the trace, reporting out-of-order timestamps and dangling node and qubit references. Each event is a plain JSON object -- carrying its type, its time and a few type-specific fields, as sketched in panel~(1) of Fig.~\ref{fig:overview} with the timestamps omitted -- so that a trace can be written with the JSON support of any programming language. For instance, traces are written without Q2NS by an independent producer, relying on the Python standard library only, and by a demo of SeQUeNCe~\cite{wu2021sequence} extended with a few lines, both available in the supporting material of this article. The trace format is specified in the GitHub repository (\texttt{doc/trace-format.md}), together with a JSON Schema of the events. The current specification defines version~2 of the format and allows a producer to extend its traces with additional optional fields or event types: unrecognized fields have no effect on the replay, while unrecognized event types are skipped and the remaining supported events are replayed normally. Such backward-compatible extensions retain the same format version, whereas incompatible changes require a new explicitly versioned format.

Q2NSViz is validated on the example traces shipped with the software and generated by Q2NS -- Bell-pair distribution, teleportation, entanglement swapping, graph-state manipulation and lossy channel -- listed in the repository: for each trace, the test suite checks that the incremental replay coincides, at every keyframe, with a from-scratch replay used as reference. Fig.~\ref{fig:graphstate} shows selected graphical views generated by the current Q2NSViz interface for the graph-state scenario also presented in~\cite{MazPeaCal-26}: while the simulator performs the measurement on the quantum state, Q2NSViz replays its effect on the entanglement graph through the reported rule~\cite{HeiEisBri-04}.

\section{Impact Overview}
Quantum-network simulators are becoming essential tools for studying protocol and network behavior, but their outputs often remain difficult to inspect without detailed knowledge of the simulator internals and debugging abilities. Q2NSViz turns finished simulation traces into portable, time-resolved visual artifacts: any Q2NS run can be archived, replayed, shared, and re-examined as a self-contained file. Users can scrub to the exact instant a protocol misbehaves -- an entanglement edge that never forms, a qubit lost in flight, a late correction -- rather than reconstructing it from interleaved log lines. 
Users can inspect the bundled traces and navigate their execution without interacting with the simulator internals or writing additional code. Together with Q2NS, Q2NSViz ships with example traces covering basic to advanced quantum communication scenarios, from teleportation to graph-state manipulation and transmission over a lossy channel, showing how the protocol events modify qubits, classical information, measurements, and entanglement-induced connectivity over time. Q2NSViz also embeds non-trivial entanglement manipulation primitives: entanglement created and consumed, swapped across repeaters, and reshaped by local measurements. This makes the tool useful for debugging, education, and demonstrations. Indeed, the browser-based predecessor of Q2NSViz supported the inspection and validation of Q2NS simulations~\cite{PeaMazCal-26}, whereas the standalone Python implementation was used to visualize entanglement manipulation in~\cite{MazPeaCal-26}, including the graph-state scenario shown in Fig.~\ref{fig:graphstate}.

\section{Conclusions}
This article presented Q2NSViz, an open-source standalone tool for replaying and inspecting quantum-network simulation traces. By combining a documented trace format, a Qt-free replay engine, and an interactive Python/PyQt6 interface, Q2NSViz turns simulator output into a time-resolved, navigable, and reproducible representation of quantum-network execution. It therefore complements Q2NS~\cite{PeaMazCal-26,MazPeaCal-26,PeaCalCac-26} not only as a visual frontend, but as a software layer for inspecting, sharing, and comparing protocol dynamics, and as a reusable basis for future Q2NS releases and compatible simulation backends.

\begin{table}[t]
  \centering
  \footnotesize
  \setlength{\tabcolsep}{2pt}
  \renewcommand{\arraystretch}{1.02}
  \begin{tabularx}{\columnwidth}{@{}>{\ttfamily}l >{\raggedright\arraybackslash}X@{}}
    \toprule
    \normalfont\bfseries Event & \normalfont\bfseries Effect on the views \\
    \midrule
    \rowcolor{qprimary!16} createNode    & \normalfont Node card placed on the canvas. \\
    \rowcolor{qprimary!16} createChannel & \normalfont Quantum or classical link drawn between two nodes. \\
    \rowcolor{qproduct!14} createQubit   & \normalfont Qubit marker appears at its host node. \\
    \rowcolor{qproduct!14} removeQubit   & \normalfont Qubit cleared from the graph, red $\times$ if lost. \\
    \rowcolor{qentangle!13} entangle     & \normalfont Toggles edge(s) among qubits on the entanglement graph. \\
    \rowcolor{qentangle!13} measure      & \normalfont Bell/Pauli measurement: qubit grayed and removed. \\
    \rowcolor{qentangle!13} graphMeasure & \normalfont Graph-state $X/Y/Z$ measurement via local complementation. \\
    \rowcolor{qquantum!15} sendQubit     & \normalfont Qubit animated along the quantum channel over $[t_0,t_1]$. \\
    \rowcolor{qclassic!24} createCbit    & \normalfont Classical-bit marker appears at its host node. \\
    \rowcolor{qclassic!24} sendCbit      & \normalfont Classical-bit marker animated along the classical channel. \\
    \rowcolor{qclassic!24} removeCbit    & \normalfont Classical bit cleared from its node. \\
    \rowcolor{qclassic!24} sendPacket    & \normalfont Classical packet animated along the channel. \\
    \rowcolor{black!8} traceText         & \normalfont Message shown as overlay and in the event-log tab. \\
    \bottomrule
  \end{tabularx}
  \caption{\raggedright Trace events by semantic category:
  \colorbox{qprimary!16}{network topology}, \colorbox{qproduct!14}{qubit lifecycle}, \colorbox{qentangle!13}{entanglement manipulation}, \colorbox{qquantum!15}{quantum communication}, \colorbox{qclassic!24}{classical communication}, and \colorbox{black!8}{annotation}.}
  \label{tab:events}
\end{table}

\subsubsection*{\small CRediT authorship contribution statement}
F.M.: Writing – original draft, Writing – review \& editing, Visualization, Validation, Software, Investigation, Methodology, Data curation. M.C.: Writing – review \& editing, Supervision, Methodology. A.S.C.: Writing – original draft, Writing – review \& editing, Supervision, Project administration, Methodology, Funding acquisition, Conceptualization.

\subsubsection*{\small Declaration of competing interest}
The authors declare that they have no known competing financial interests or personal relationships that could have appeared to influence the work reported in this paper.

\subsubsection*{\small Acknowledgements}
This work has been funded by the European Union under Horizon Europe ERC-CoG grant QNattyNet, n.101169850. Views and opinions expressed are however those of the author(s) only and do not necessarily reflect those of the European Union or the European Research Council Executive Agency. Neither the European Union nor the granting authority can be held responsible for them.

\subsubsection*{\small Data availability}
Q2NSViz is openly available on GitHub and
archived on Zenodo at \url{https://doi.org/10.5281/zenodo.21216676}. 
The supporting material of this article is also available not only on GitHub but also on Zenodo at \url{doi.org/10.5281/zenodo.23940911}.

\setlength{\bibsep}{0pt plus 0.3ex}
\bibliographystyle{IEEEtranN}
\bibliography{references}

\end{document}